\documentclass{PoS}
\usepackage{amsmath}

\def\slashb#1{\not\!\!#1}

\title{Lattice QCD study for relation between\\
confinement and chiral symmetry breaking\\
on temporally odd-number lattice}

\ShortTitle{Lattice QCD study for relation between confinement \& CSB on temporally odd lattice}

\author{\speaker{Takahiro M. Doi}, Hideo Suganuma \\
        Department of Physics \& Division of Physics and Astronomy, 
Graduate School of Science, \\
Kyoto University, 
Kitashirakawaoiwake, Sakyo, Kyoto 606-8502, Japan\\
        E-mail: \email{doi@ruby.scphys.kyoto-u.ac.jp}}

\author{Takumi Iritani \\
High Energy Accelerator Research Organization (KEK), 
Tsukuba, Ibaraki 305-0801, Japan}

\abstract{
We investigate the contribution from each Dirac modes to the Polyakov loop 
based on a gauge-invariant analytical relation 
connecting the Polyakov loop and the Dirac modes on a temporally odd-number 
lattice, where the temporal lattice size is odd, 
with the normal (nontwisted) periodic boundary condition. 
The dumping factor in the relation plays crucial role 
for the negligible contribution of low-lying Dirac modes to the Polyakov loop. 
The zero-value of the Polyakov loop in the confinement phase is due to the 
``positive/negative symmetry'' of 
the Dirac-mode contribution to the Polyakov loop. 
In the deconfinement phase, there is no such symmetry. 

}
\FullConference{XV International Conference on Hadron Spectroscopy-Hadron 2013\\
		4-8 November 2013\\
		Nara, Japan }

\begin{document}

\section{Introduction}
Color confinement and chiral symmetry breaking have been investigated 
as interesting non-perturbative phenomena in low-energy QCD 
in many analytical and numerical studies. 
However, their properties are not sufficiently understood directly from QCD. 
The Polyakov loop is an order parameter for quark confinement \cite{Rothe}. 
At the quenched level, the Polyakov loop is the exact order parameter for 
quark confinement, and its expectation value is zero in confinement phase 
and is nonzero in deconfinement phase. 
Also, its fluctuation is recently found to be important in the QCD phase transition \cite{Redlich}.
As for the chiral symmetry, 
low-lying Dirac modes are essential for chiral symmetry breaking in QCD, 
according to the Banks-Casher relation \cite{BC}.

Not only the properties of confinement and chiral symmetry breaking in QCD 
but also their relation is an interesting challenging subject \cite{Suganuma,G06BGH07}. 
From many analytical and numerical studies, 
it is suggested that confinement and chiral symmetry breaking are strongly correlated 
\cite{Miyamura,Karsch}.
However, we showed analytically and numerically
%an analytical relation between the Polyakov loop and Dirac modes on temporally odd-number lattice, and 
that low-lying Dirac modes have little contribution to the Polyakov loop 
and that there is no one-to-one correspondence 
between confinement and chiral symmetry breaking in QCD \cite{GIS,SDI,DSI}. 

%By removing QCD monopoles in the maximally Abelian gauge, 
%both confinement and chiral symmetry breaking are lost 
%in lattice QCD \cite{Miyamura}.
%The transition temperatures of deconfinement and chiral restoration 
%are almost same in finite temperature QCD \cite{Karsch}. 
%From these facts, it is suggested that 
%confinement and chiral symmetry breaking are strongly correlated.
%In recent lattice-QCD numerical studies, however, 
%it is found that confinement properties 
%do not change by removing low-lying Dirac modes from the QCD vacuum, 
%which means no one-to-one correspondence between 
%confinement and chiral symmetry breaking in QCD \cite{GIS}.

In this study, we discuss the relation between confinement and chiral symmetry breaking 
based on an analytical relation between the Polyakov loop and Dirac modes on temporally odd-number lattice, 
with the normal (nontwisted) periodic boundary condition \cite{SDI,DSI}. 
We investigate each Dirac-mode contribution to the Polyakov loop in both confinement and deconfinement phases.

\section{Dirac modes in lattice QCD}
In this section, we review the Dirac operator, 
its eigenvalues and its eigenmodes (Dirac modes) in ${\rm SU}(N_{\rm c})$ 
lattice QCD \cite{GIS}.
We use a standard square lattice with spacing $a$, and the notation of 
sites $s=(s_1, s_2, s_3, s_4) \ (s_\mu=1,2,\cdots,N_\mu)$, 
and link-variables $U_\mu(s)={\rm e}^{iagA_\mu(s)}$ 
with gauge fields $A_\mu(s) \in su(N_c)$ and gauge coupling $g$.
In lattice QCD, the Dirac operator $\slashb{D}=\gamma_\mu D_\mu$ 
is given by
\begin{eqnarray}
 \slashb{D}_{s,s'} 
      = \frac{1}{2a} \sum_{\mu=1}^4 \gamma_\mu 
\left[ U_\mu(s) \delta_{s+\hat{\mu},s'}
        - U_{-\mu}(s) \delta_{s-\hat{\mu},s'} \right], \label{Eq:DiracOp}
\end{eqnarray}
with $U_{-\mu}(s)\equiv U^\dagger_\mu(s-\hat\mu)$.
Here, $\hat\mu$ is the unit vector in direction $\mu$ in the lattice unit.
In this paper, we define all the $\gamma$-matrices to be hermite as 
$\gamma_\mu^\dagger=\gamma_\mu$.
Since the Dirac operator is anti-hermite 
in this definition of $\gamma_\mu$, 
the Dirac eigenvalue equation is expressed as
\begin{eqnarray}
\slashb{D}|n\rangle =i\lambda_n|n \rangle
\end{eqnarray}
with the Dirac eigenvalue $i\lambda_n$ ($\lambda_n \in {\bf R}$) 
and the Dirac eigenstate $|n \rangle$.
%Note that the chiral partner $\gamma_5|n\rangle$ is also 
%an eigenstate with the eigenvalue $-i\lambda_n$.
%Using the Dirac eigenfunction $\psi_n(s)\equiv\langle s|n \rangle $, 
%the explicit form for the Dirac eigenvalue equation is written by 
%\begin{eqnarray}
% \frac{1}{2a}& \sum_{\mu=1}^4 \gamma_\mu
%[U_\mu(s)\psi_n(s+\hat \mu)-U_{-\mu}(s)\psi_n(s-\hat \mu)]=i\lambda_n \psi_n(s). \label{Eq:eigenEqExplicit}
%\end{eqnarray}

\vspace{-0.45cm}

\section{An analytical relation between the Polyakov loop 
and Dirac modes on temporally odd-number lattice}

We consider a temporally odd-number lattice, 
where the temporal lattice size $N_4$ is odd, 
with the normal (nontwisted) periodic boundary condition 
in both temporal and spatial directions. 
The spatial lattice size $N_{1 \sim 3} (> N_4)$ is taken to be even.
Using Elitzur's theorem, 
we derive a relation connecting the Polyakov loop 
and the Dirac modes \cite{SDI,DSI},
\begin{eqnarray}
\langle L_P \rangle=\frac{(2ai)^{N_4-1}}{12V}
\sum_n\lambda_n^{N_4-1}\langle n|\hat{U}_4| n \rangle,  \label{Eq:RelOrig}
\end{eqnarray}
where the link-variable operator $\hat{U}_{\pm\mu}$ is defined 
by the matrix element 
\begin{eqnarray}
\langle 
s | \hat{U}_{\pm\mu} |s' \rangle=U_{\pm\mu}(s)\delta_{s\pm\hat{\mu},s'}.
\end{eqnarray}
%The derivation of the relation is shown in our previous paper \cite{SDI,DSI}.
In this derivation, we use the fact that 
any gauge-invariant quantity cannot be composed by the product of 
odd-number $N_4$ link-variables \cite{SDI,DSI}, 
except for the Polyakov loop. 
This is a Dirac spectral representation of the Polyakov loop, 
and is valid on the temporally odd-number lattice. 
Using this relation  (\ref{Eq:RelOrig}), 
we can investigate each Dirac-mode contribution 
to the Polyakov loop individually 
and discuss the relation between confinement 
and chiral symmetry breaking in QCD.

\section{Modified KS formalism for temporally odd-number lattice}
The Dirac operator $\slashb{D}$ has a large dimension of 
$(4 \times N_{\rm c}\times V)^2$, so that 
the numerical cost for solving the Dirac eigenvalue equation is quite huge.
This numerical cost can be partially reduced 
using the Kogut-Susskind (KS) formalism \cite{Rothe,GIS,KS}.
However, the original KS formalism can be applied only to the ``even lattice'' 
where all the lattice sizes $N_\mu$ are even number. 
In this section, we show brief introduction of the modified KS formalism \cite{DSI}
applicable to the odd-number lattice. 
%Using the modified KS formalism, we can reduce the numerical cost 
%in the case of the temporally odd-number lattice.

We consider the temporally odd-number lattice, 
%with all the spatial lattice size being even
and introduce a matrix 
\begin{eqnarray}
 M(s)\equiv\gamma_1^{s_1}\gamma_2^{s_2}\gamma_3^{s_3}\gamma_4^{s_1+s_2+s_3}. 
\label{Eq:M}
\end{eqnarray}
%As a remarkable feature, the requirement from 
%the periodic boundary condition is satisfied 
%on the temporally odd-number lattice: $M(s+N_\mu\hat{\mu})=M(s) \ (\mu=1,2,3,4).$
Using the matrix $M(s)$ and taking the Dirac representation, 
we can spin-diagonalize the Dirac operator $\slashb{D}$ in the case of 
the temporally odd-number lattice,
\begin{eqnarray}
 \sum_\mu M^\dagger(s) \gamma_\mu D_\mu M(s+ \hat \mu) = {\rm diag}(\eta_\mu D_\mu,\eta_\mu D_\mu,-\eta_\mu D_\mu,-\eta_\mu D_\mu), \label{Eq:MDiracM}
\end{eqnarray}
where $\eta_\mu D_\mu$ is the KS Dirac operator given by
\begin{eqnarray}
(\eta_\mu D_\mu)_{ss'}=\frac{1}{2a}\sum_{\mu=1}^{4}\eta_\mu(s)\left[U_\mu(s)\delta_{s+\hat{\mu},s'}-U_{-\mu}(s)\delta_{s-\hat{\mu},s'}\right]. \label{Eq:KSDiracOp}
\end{eqnarray}
Here, $\eta_\mu(s)$ is the staggered phase:
$\eta_1(s)\equiv 1, \ \ \eta_\mu(s)\equiv (-1)^{s_1+\cdots+s_{\mu-1}} 
\ (\mu \geq 2) \label{Eq:eta}$.
%Then, for each $\lambda_n$, 
%two positive modes and two negative modes appear 
%relating to the spinor structure on the temporally odd-number lattice.
%(Note also that the chiral partner $\gamma_5 |n \rangle$ 
%gives an eigenmode with the eigenvalue $-i\lambda_n$.)
Thus, all the eigenvalues $i\lambda_n$ can be obtained by 
solving the reduced Dirac eigenvalue equation, 
\begin{eqnarray}
\eta_\mu D_\mu|n) =i\lambda_n|n ). \label{Eq:KSEigenEqOpOdd}
\end{eqnarray}

\section{Numerical analysis for each Dirac-mode contribution to the Polyakov loop}

Using the modified KS formalism, Eq.(\ref{Eq:RelOrig}) is rewritten as
\begin{eqnarray}
\langle L_P \rangle =\frac{(2ai)^{N_4-1}}{3V}
\sum_n\lambda_n^{N_4-1}( n|\hat{U}_4| n ). \label{Eq:RelKS}
\end{eqnarray}
Note that the (modified) KS formalism is an exact method for diagonalizing 
the Dirac operator and is not an approximation, so that 
Eqs.(\ref{Eq:RelOrig}) and (\ref{Eq:RelKS}) are completely equivalent.

We numerically calculate each Dirac-mode contribution to the Polyakov loop, 
i.e., the matrix elements $( n|\hat{U}_4| n )$ and $\lambda_n^{N_4-1}( n|\hat{U}_4| n )$ 
in the sum of the relation (\ref{Eq:RelKS}). 
We perform SU(3) lattice QCD Monte Carlo simulations with 
the standard plaquette action at the quenched level 
in both cases of confinement and deconfinement phases.
For the confinement phase, we use $10^3\times 5$ lattice with 
$\beta\equiv2N_{\rm c}/g^2=5.6$ (i.e., $a\simeq 0.25~{\rm fm}$), 
corresponding to $T\equiv1/(N_4a)\simeq160~{\rm MeV}$.
For the deconfinement phase, we use $10^3\times 3$ lattice with $\beta=5.7$ 
(i.e., $a\simeq 0.20~{\rm fm}$), corresponding to 
$T\equiv1/(N_4a)\simeq 330~{\rm MeV}$.
For each phase, 
we use 20 gauge configurations, which are taken every 500 sweeps 
after the thermalization of 5,000 sweeps.

As the numerical result, 
we find that the relation (\ref{Eq:RelKS}) is almost exact and 
low-lying Dirac modes have little contribution to the Polyakov loop 
for each gauge configuration in both confinement and deconfinement phases \cite{DSI}. Thus, we can discuss each Dirac-mode contribution to the Polyakov loop 
even for one gauge configuration. 

In the confinement phase, we show in Fig.1 
each Dirac-mode contribution to the Polyakov loop 
$\lambda_n^{N_4-1}( n|\hat{U}_4| n )$
as the function of Dirac eigenvalue $\lambda_n$. 
From Fig. 1, we can confirm that low-lying Dirac modes have little contribution to the Polyakov loop. 
All the sum of these quantities $\lambda_n^{N_4-1}( n|\hat{U}_4| n )$ is zero, 
which leads to the vanishing Polyakov loop in the confinement phase.
As a remarkable fact, the zero value of the Polyakov loop is 
due to the ``positive/negative symmetry'' of 
real and imaginary parts of $\lambda_n^{N_4-1}( n|\hat{U}_4| n )$, 
as shown in Fig.1. 
\begin{figure}[h]
\begin{center}
\includegraphics[scale=0.4]{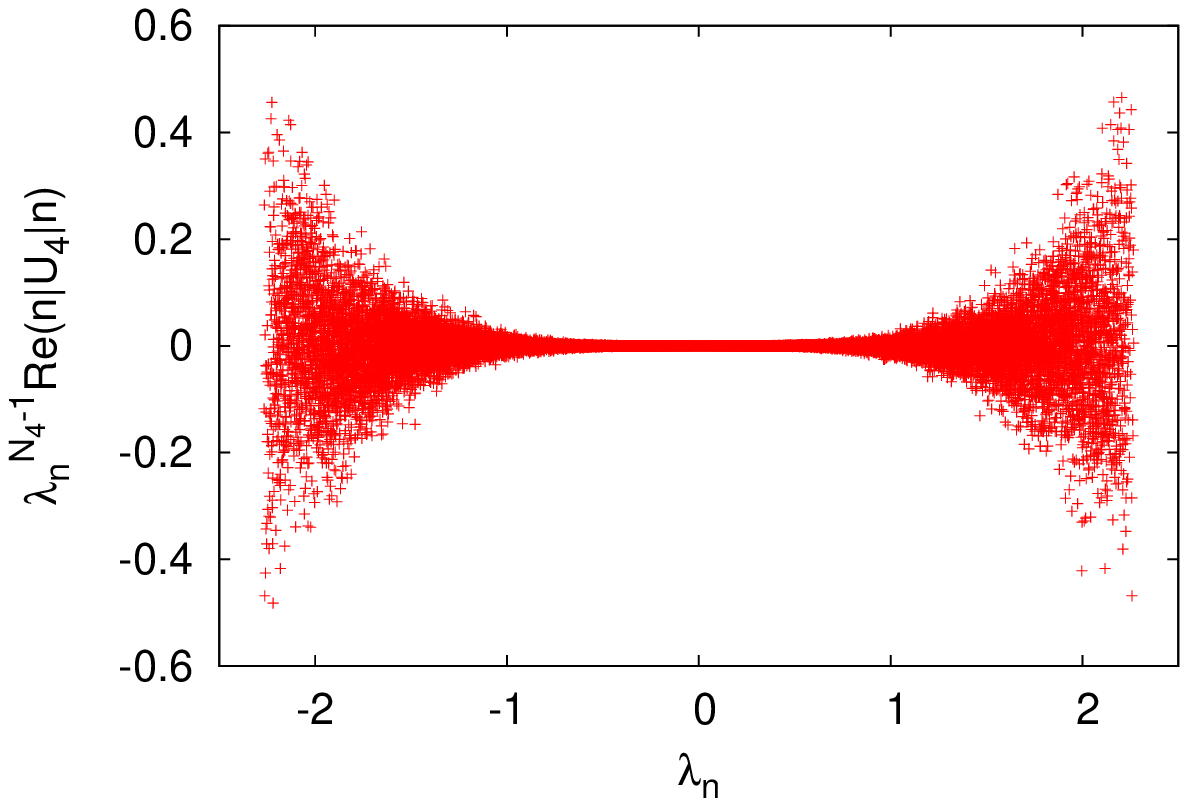}
\hspace{0.4cm} \includegraphics[scale=0.4]{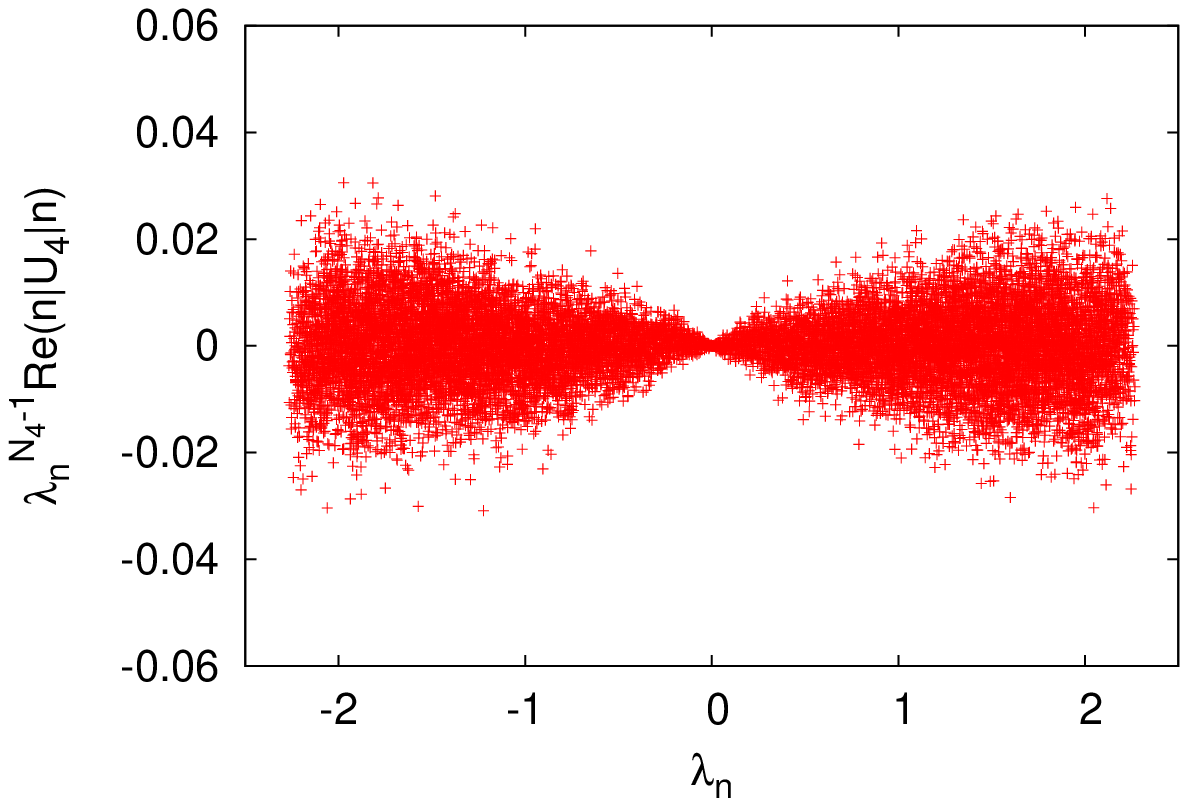}
\vspace{-0.25cm}
\caption{
Lattice QCD results for each Dirac-mode contribution to the Polyakov loop, 
$\lambda_n^{N_4-1}( n|\hat{U}_4| n )$, 
as the function of the Dirac eigenvalue $\lambda_n$ in the lattice unit, 
on $10^3\times5$ with $\beta=5.6$, for one gauge configuration 
in the confinement phase. 
The left figure shows the real part and the right figure the imaginary part.
}
\end{center}
\end{figure}

In the deconfinement phase, we show in Fig.2 
the real part of the matrix elements $( n|\hat{U}_4| n )$ and 
each Dirac-mode contribution to the Polyakov loop, 
$\lambda_n^{N_4-1}( n|\hat{U}_4| n )$, 
as the function of Dirac eigenvalue $\lambda_n$. 
In this configuration, the expectation value of the Polyakov loop is real, 
and the behavior of the imaginary part of these quantities is similar 
to that in the confinement phase. 
There is no ``positive/negative symmetry'' 
in the DIrac-mode distribution of ${\rm Re} ( n|\hat{U}_4| n )$, 
and all the sum of these quantities 
$\lambda_n^{N_4-1}{\rm Re}( n|\hat{U}_4| n )$ is nonzero, 
which gives a nonzero value of 
the Polyakov loop in the deconfinement phase. 
The signs of the contribution from infrared Dirac modes 
and ultraviolet Dirac modes are different. 
%This was suggested in previous works \cite{G06BGH07}.
Although the matrix elements ${\rm Re}( n|\hat{U}_4| n )$ 
have a peak in the small Dirac-mode region, 
the contribution from this region to the Polyakov loop 
is very small, 
because of the dumping factor $\lambda_n^{N_4-1}$. 
Thus, the factor $\lambda_n^{N_4-1}$ plays a crucial role 
in RHS of Eq.(\ref{Eq:RelKS}).

\begin{figure}[h]
\begin{center}
\includegraphics[scale=0.4]{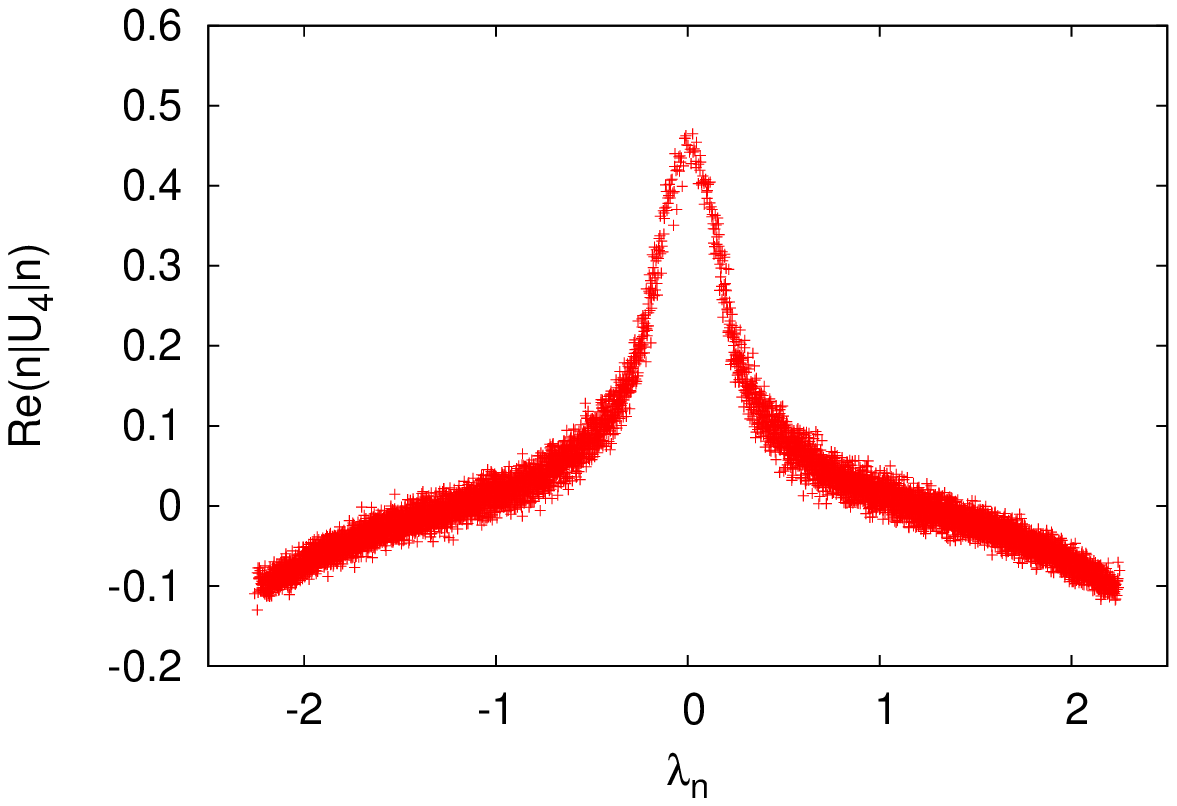}
\hspace{0.4cm} \includegraphics[scale=0.4]{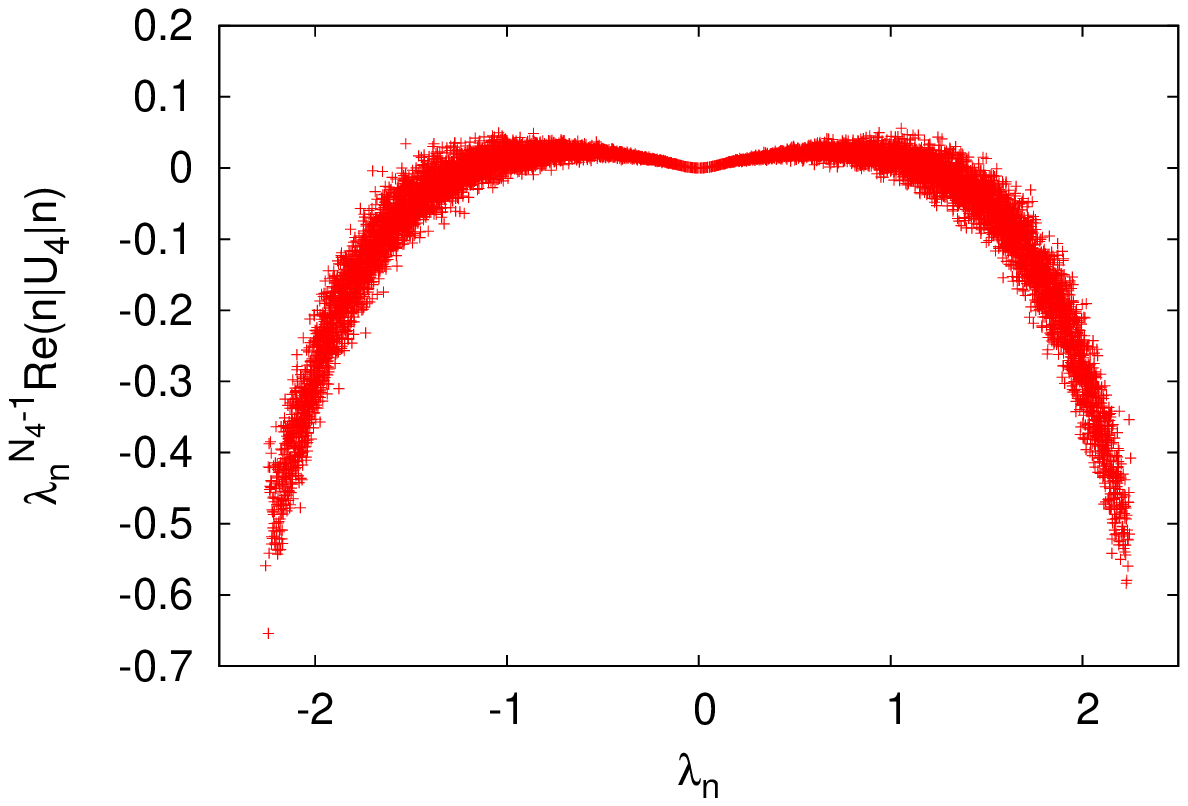}
\vspace{-0.25cm}
\caption{
Lattice QCD results for the real part of the matrix element 
$\lambda_n^{N_4-1}{\rm Re}( n|\hat{U}_4| n )$ 
and each Dirac-mode contribution to the Polyakov loop, 
$\lambda_n^{N_4-1}{\rm Re}( n|\hat{U}_4| n )$, 
as the function of the Dirac eigenvalue $\lambda_n$ 
in the lattice unit, on $10^3\times3$ with $\beta=5.7$, 
for one gauge configuration in the deconfinement phase. 
}
\vspace{-0.55cm}
\end{center}
\end{figure}

\section{Summary and concluding remarks}

\vspace{-0.13cm}

In this study we have investigated 
each Dirac-mode contribution to the Polyakov loop 
based on the relation connecting the Polyakov loop and the Dirac modes 
%in both confinement and deconfinement phases
on temporally odd-number lattice, 
with the normal (nontwisted) periodic boundary condition. 
In both confinement and deconfinement phases, 
low-lying Dirac modes have little contribution to the Polyakov loop 
because of the dumping factor 
$\lambda_n^{N_4-1}$ in RHS of Eq.(\ref{Eq:RelKS}). 
Also, the zero-value of the Polyakov loop in confinement phase is 
due to the ``positive/negative symmetry'' of the 
Dirac-mode matrix elements $( n|\hat{U}_4| n )$. 
In the deconfinement phase, there is no such symmetry. 

\vspace{-0.13cm}

\section*{Acknowledgements}

\vspace{-0.13cm}

H.S. and T.I. are supported in part by the Grant for Scientific Research 
[(C) No.23540306, 
%Priority Areas ``New Hadrons'' 
E01:21105006, No.21674002] from the Ministry of Education, 
Science and Technology of Japan.  
The lattice QCD calculation has been done on NEC-SX8R at Osaka University.

\end{document}